# Microgrid Operation Control with Adaptable Droop Gains

Edgar Diego Gomez Anccas
*Chair of Electrical Power Systems*
*Helmut Schmidt University*
Hamburg, Germany
diego.gomez@hsu-hh.de

Christian A. Hans
*ASN Group*
*University of Kassel*
Kassel, Germany
hans@uni-kassel.de

Detlef Schulz
*Chair of Electrical Power Systems*
*Helmut Schmidt University*
Hamburg, Germany
detlef.schulz@hsu-hh.de

*Abstract*—Modern low-carbon power systems come with many challenges, such as increased inverter penetration and increased uncertainty from renewable sources and loads. In this context, the microgrid concept is a promising approach, which is based on a segmentation of the grid into independent smaller cells that can run either in grid-connected or standalone mode.In microgrids, droop control is widely used for primary control. It enables proportional power sharing, depending on the droop gains. Operation control schemes considering droop control often assume fixed droop gains. However, using adaptive droop gains for grid-forming units allow to shape power sharing in presence of fluctuations, enhancing flexibility while maintaining a safe microgrid operation, particularly under uncertainty. This work introduces a bilinear formulation for microgrid operation control that finds optimal power setpoints and droop gains on a timescale of minutes by solving a finite horizon optimization problem. In detail, a robust minmax model predictive control scheme is designed for a standalone microgrid, comprising a fuel cell, a photovoltaic system and an energy storage. Closed-loop simulations are performed with and without variable droop gains. The results show an increase in renewable utilization of up to 7.5 % while reducing the power output of the fuel cell by 6 %, when allowing variable droop gains.

## I. Introduction

Modern energy systems are transitioning towards an increasingly decentralized structure that includes a large number of renewable sources across different voltage levels. This amplifies the uncertainty in the grid operation due to the volatile renewable infeed [1]. The transition can be addressed using the microgrid (MG) approach of segmenting the grid into smaller cells. Each can be operated in either grid-connected or standalone mode. Standalone operation is particularly challenging for MGs as they are isolated from the grid and required to balance generation and consumption locally. In this context, operation control provides optimal power setpoints for the lower control layers, which are derived by minimizing an objective function subject to constraints that model the MG. Key objectives are a reliable MG operation and a maximization of economic benefits, e.g., by maximizing renewable utilization.

In this context, model predictive control (MPC) is a promising approach. It takes load and renewable infeed forecasts and initial conditions of states as inputs and solves optimization problems in a receding horizon fashion. Depending on the application, different MPC types are used [2]: Certainty equivalence MPC, where the nominal forecast is assumed to be true, is widely used in operation control. Minmax-MPC is an alternative, where bounded uncertain variables are assumed. Although the solution yields conservative setpoints [3], it ensures safe operation by explicitly modelling uncertainties, using forecast intervals of load and renewable infeed. In MGs, the uncertainty also affects power and energy values which in MPC formulations also become elements of bounded intervals. In MGs with inverter-based sources only, grid-forming control is required for standalone operation. In this context, droop control is widely used. For the operation of parallel sources, droop gains determine the share of fluctuations that each source covers. To ensure constraint satisfaction of power and energy, droop control should be included in optimization schemes that consider uncertainties. In MG optimization schemes these gains are often fixed through low layer control design.

However, using time-varying droop gains for grid-forming units allow us to shape the sharing of fluctuations, which enables more flexibility while maintaining a safe MG operation. So far, only a small number of publications consider dynamic droop gains in MG optimization schemes. In [4], droop gains are adapted based on wind turbine reserves. The method focuses on primary control and omits MPC and robust forecast intervals. In [5], the MPC focuses on optimizing droop gains and voltage setpoints for stability, without targeting an operational schedule. In [6], a linear power-sharing approach defines droop gains based on cost-cases to reduce generation costs, but lacks MPC and operates at a lower control layer. In [7], mixed-integer conic programming is used within a stochastic optimization framework to derive piecewise-affine power-dependent droop gains offline. In [8], adaptive droop gains minimize power loss and costs in a multi-layer control scheme, however, updates are based on grid conditions or time intervals, and do not consider uncertainty. Similarly, [9] varies droop gains to reduce transmission losses, but lacks MPC and robust forecast intervals. Lastly, [10] introduces an efficiency based MG model for parameter optimization to improve operation efficiency under varying load, but lacks robust forecasts and MPC. Although existing publications provide valuable contributions,most of them do not address operation control directly and omit uncertainty.

This work extends [11] by introducing an operation control formulation with adaptive droop gains for standalone MGs, represented as bilinear constraints. They enable the distribution of power and shape uncertainty intervals of grid-forming units within the MG in a minmax MPC scheme. In a numerical case study, the benefits of the adaptive configuration, compared to a fixed one, are evaluated through closed-loop simulations.

TABLE I
MODEL-SPECIFIC VARIABLES

| Symbol | Explanation | Domain | Unit |
|---|---|---|---|
| $x$ | Energy of storage unit (state) | $\mathbb{R}_{>0}$ | pu h |
| $u_\text{F}$ | Setpoint input of FC unit | $\mathbb{R}_{>0}$ | pu |
| $u_\text{B}$ | Setpoint input of storage unit | $\mathbb{R}$ | pu |
| $u_\text{PV}$ | Setpoint input of PV unit | $\mathbb{R}_{\geq 0}$ | pu |
| $u$ | Setpoint inputs of all unit | $\mathbb{R}^3$ | pu |
| $\delta_\text{F}$ | Boolean input of FC unit | $\mathbb{B}$ | — |
| $\chi_\text{F}, \chi_\text{B}$ | Droop gain of FC / storage unit | $\mathbb{R}_{>0}$ | — |
| $v$ | Vector of all control input | $\mathbb{R}^6$ | pu |
| $w_\text{PV}, w_\text{L}$ | Uncertain available PV power / Load | $\mathbb{R}_{\geq 0}, \mathbb{R}_{\leq 0}$ | pu |
| $w$ | Vector of all uncertain input | $\mathbb{R}^2$ | pu |
| $p_\text{F}$ | Power of FC unit | $\mathbb{R}_{>0}$ | pu |
| $p_\text{B}$ | Power of storage unit | $\mathbb{R}$ | pu |
| $p_\text{PV}$ | Power of PV unit | $\mathbb{R}_{\geq 0}$ | pu |
| $p$ | Power of all unit | $\mathbb{R}^3$ | pu |
| $\delta_\text{PV}$ | Boolean auxiliary variables | $\mathbb{B}$ | — |
| $z$ | Vector of all auxiliary variables | $\mathbb{R}^5$ | — |

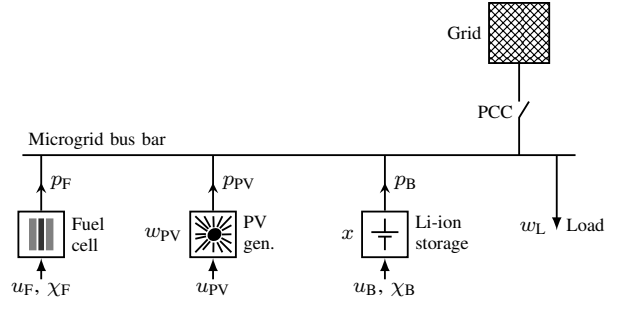

Fig. 1. Microgrid under investigation. Illustration based on [13].

This paper is structured as follows. Section II introduces the minmax MPC scheme, including the MG model with bilinear constraints. Section III presents closed-loop simulations of the two configurations. Finally, Section IV concludes the paper.

*A. Notation*

The sets of nonnegative integers, real numbers, positive real numbers, nonnegative real numbers and negative real numbers are $\mathbb{N}_0$, $\mathbb{R}$, $\mathbb{R}_{>0}$, $\mathbb{R}_{\geq 0}$ and $\mathbb{R}_{<0}$, respectively. The set of Booleans is $\mathbb{B} = \{0, 1\}$. A forecast of variable $a$ at time $k \in \mathbb{N}_0$ for a future timestep $k + j \in \mathbb{N}$ is described by $a(k+j|k)$. The vector with elements $a_i, \ldots, a_N$ is described by $[a_i]_{i=1}^N = [a_i \cdots a_N]^\top$ with $N \in \mathbb{N}$. The set of nonnegative integers in the interval $[a, b] \subset \mathbb{N}_0$ is $\mathbb{N}_{[a,b]}$.

## II. MINMAX MPC SCHEME

As uncertainties are part of real systems, they need to be considered. An extension to "traditional" certainty-equivalence MPC is minmax MPC [12], where robust control actions are found by solving optimization problems that consider uncertainty and minimize the worst-case cost. Based on monotonicity properties, we model the uncertain input, which is element of a closed set, in the form of upper and lower bounds.

In this work, as shown in Fig. 1, the system under investigation is a low-voltage AC MG composed of four units connected to a point of common coupling (PCC) and operated in standalone mode: A fuel cell (FC) and a battery system are droop-controlled, forming the grid. A curtailable photovoltaic (PV) system and a symmetrical three phase load complete the MG. The model variables are listed in Table I.

*A. Microgrid Model*

In our model, which is based on [13], the constraints are either monotonically increasing or monotonically decreasing in the bounded uncertain input $w(k) \in [w^\text{low}(k), w^\text{high}(k)] \subset \mathbb{R}^2$. To account for this full inverval, it is therefore sufficient to consider only the extreme cases $w^\text{low}(k)$ and $w^\text{high}(k)$. Hence, constraints only need to be defined for $\sigma \in \{\text{high}, \text{low}\}$. The PV unit's power and setpoint limits are given by

$$p_\text{PVmin} \leq p_\text{PV}^\sigma(k) \leq p_\text{PVmax}, \quad (1a)$$
$$p_\text{PVmin} \leq u_\text{PV}(k) \leq p_\text{PVmax}, \quad (1b)$$

with $p_\text{PVmax} \in \mathbb{R}_{>0}$ and $p_\text{PVmin} \in \mathbb{R}_{\geq 0}$. Curtailment, through the setpoint, is modelled by $p_\text{PV}^\sigma(k) = \min(u_\text{PV}(k), w_\text{PV}^\sigma(k))$, or equivalently using big-M reformulations [14] by

$$u_\text{PV}(k) - M_\text{PV} \cdot \delta_\text{PV}^\sigma(k) \leq p_\text{PV}^\sigma(k) \leq u_\text{PV}(k), \quad (2a)$$
$$w_\text{PV}^\sigma(k) + m_\text{PV} \cdot (1 - \delta_\text{PV}^\sigma(k)) \leq p_\text{PV}^\sigma(k) \leq w_\text{PV}^\sigma(k). \quad (2b)$$

The battery state dynamics and limits are

$$x^\sigma(k+1) = x(k) - \Delta T \cdot p_\text{B}^\sigma(k), \quad (3a)$$
$$x_\text{min} \leq x^\sigma(k) \leq x_\text{max}, \quad x(0) = x_0, \quad (3b)$$

with $\Delta T$ is the sampling time and $x_\text{min}, x_\text{max} \in \mathbb{R}_{>0}$. The limits of $p_\text{B}$ and $u_\text{B}$ are $p_\text{Bmax} \in \mathbb{R}_{>0}$ and $p_\text{Bmin} \in \mathbb{R}_{<0}$, i.e.,

$$p_\text{Bmin} \leq p_\text{B}^\sigma(k) \leq p_\text{Bmax}, \quad (4a)$$
$$p_\text{Bmin} \leq u_\text{B}(k) \leq p_\text{Bmax}. \quad (4b)$$

The FC can be on or off, which is represented by $\delta_\text{F}(k)$ in

$$\delta_\text{F}(k) p_\text{Fmin} \leq p_\text{F}^\sigma(k) \leq p_\text{Fmax} \delta_\text{F}(k), \quad (5a)$$
$$\delta_\text{F}(k) p_\text{Fmin} \leq u_\text{F}(k) \leq p_\text{Fmax} \delta_\text{F}(k). \quad (5b)$$

Power balance is ensured by

$$p_\text{F}^\sigma(k) + p_\text{B}^\sigma(k) + p_\text{PV}^\sigma(k) = w_\text{L}^\sigma(k). \quad (6a)$$

*B. Power Sharing with Adaptive Droop Gains*

The droop control law in continuous time for unit $i$ reads

$$\dot{\theta}_i(t) = \omega_i(t) = \omega^\text{d} - \chi_i(p_i^\text{m}(t) - u_i(t)), \quad (7a)$$
$$\tau_i \dot{p}_i^\text{m}(t) = -p_i^\text{m}(t) - p_i(t), \quad (7b)$$

where $\theta_i, \omega_i, \omega^\text{d} \in \mathbb{R}$ and $p_i^\text{m}, \tau_i \in \mathbb{R}_{>0}$ denote phase angle, frequency, reference frequency, measured power and low pass filter time constant, respectively. If the MG exhibits synchronized motion [15], then $\dot{\theta}_i(t) - \dot{\theta}_j(t) = 0$ and $\dot{p}_i^m(t) = 0$, i.e, $p_i^m(t) = p_i(t)$. Thus, $\omega_i(t) = \omega_j(t) = \omega(t)$ for grid-forming units $i, j$ in the system. In this case, proportional power sharing among the grid forming units in the grid is achieved, which in discrete time reads

$$(p_i(k) - u_i(k))\chi_i = (p_j(k) - u_j(k))\chi_j. \quad (8)$$

Thus, deviations of $p_i$ from $u_i$ are shared in a predefined proportional manner and come with frequency offsets $\mu(k) = \omega^d - \omega(k)$. Taking the FCs on/off condition into account, power sharing of the grid forming units can be described by

$$(p_B^\sigma(k) - u_B(k))\chi_B = \mu^\sigma(k), \quad (9a)$$
$$(p_F^\sigma(k) - u_F(k))\chi_F = \mu^\sigma(k)\delta_F(k). \quad (9b)$$

With this notation, uncertainty affects all grid-forming units in a predefined manner. Selective power sharing is introduced by treating the gains as decision variables $\chi_F(k) \in \mathbb{R}$, $\chi_B(k) \in \mathbb{R}$. The resulting bilinear constraints for the battery are

$$(p_B^\sigma(k) - u_B(k))\chi_B(k) = \mu^\sigma(k), \quad (10)$$

and using a big-M reformulation to model the FC,

$$(p_F^\sigma(k) - u_F(k))\chi_F(k) \leq M_F \delta_F(k), \quad (11a)$$
$$(p_F^\sigma(k) - u_F(k))\chi_F(k) \geq m_F \delta_F(k), \quad (11b)$$
$$(p_F^\sigma(k) - u_F(k))\chi_F(k) \leq \mu_F^\sigma(k) - m_F(1 - \delta_F(k)), \quad (11c)$$
$$(p_F^\sigma(k) - u_F(k))\chi_F(k) \geq \mu_F^\sigma(k) - M_F(1 - \delta_F(k)). \quad (11d)$$

To avoid gains that entail large frequency deviations or jeopardise stability [16], limits are introduced, i.e,

$$\chi_{F\min} \leq \chi_F(k) \leq \chi_{F\max}, \quad \chi_{F\min}, \chi_{F\max} \in \mathbb{R} > 0, \quad (12a)$$
$$\chi_{B\min} \leq \chi_B(k) \leq \chi_{B\max}, \quad \chi_{B\min}, \chi_{B\max} \in \mathbb{R} > 0,. \quad (12b)$$

Additionally, a limitation on frequency deviations of the form

$$\mu_{\min} \leq \mu^\sigma(k) \leq \mu_{\max} \quad (13)$$

is employed to consider power quality requirements and reduce stress for a potential secondary control layer.

### C. Costs

The cost function is only composed of parts that are implicitly convex in $w(k)$. Therefore, it is sufficient to only consider the boundary cases. FC running costs are defined as

$$l_{Fru}^\sigma(k) = c_{Fru} \cdot \delta_F(k) + c'_{Fru} \cdot p_F^\sigma(k), \quad (14)$$

with $c_{Fru} \in \mathbb{R}_{>0}$ and $c'_{Fru} \in \mathbb{R}_{>0}$. Since on/off switching is detrimental to FC lifetime [17], it is penalized by

$$l_{Fsw}(k) = c_{Fsw} \cdot (\delta_F(k) - \delta_F(k-1))^2, \quad (15)$$

with $c_{Fsw} \in \mathbb{R}_{>0}$. The use of PV energy is incentivized by penalizing power curtailment through

$$l_{PV}^\sigma = c''_{PV}(p_{PV\max}(k) - p_{PV}^\sigma(k))^2, \quad (16)$$

with $c''_{PV} \in \mathbb{R}_{>0}$. The losses during battery operation are captured by quadratic costs, weighted by $c''_B \in \mathbb{R}_{>0}$, i.e.,

$$l_B^\sigma(k) = c''_B(p_B^\sigma(k))^2. \quad (17)$$

Hence, the overall cost can be formed as

$$l^\sigma(k) = l_{Fru}^\sigma(k) + l_{Fsw}(k) + l_B^\sigma(k) + l_{PV}^\sigma(k) \quad (18)$$

The aim of the MPC is minimizing the cost of the worst case scenario. Given the convexity in $w(k)$ of the cost function at each timestep $k$ and with the auxiliary variable $t(k)$ the objective can be expressed along the lines of [3] by

$$\max_{\sigma \in \{\text{high,low}\}} l^\sigma(k) = \min_{\substack{t(k) \geq l^{\text{high}}(k) \\ t(k) \geq l^{\text{low}}(k)}} t(k). \quad (19)$$

The cost over the prediction horizon $J \in \mathbb{N}$ is then

$$l = \sum_{j=0}^{J-1} \min t(k+j|k) \cdot \gamma^{j+1}. \quad (20)$$

### D. MPC Formulation

The MPC problem combines the worst-case costs with MG model constraints and optimizes solely for the robust forecast intervals' boundary cases (see Fig. 2). This is sufficient since values within the interval, are also covered due to the monotonicity in $w(k)$ of the constraints and the convexity in $w(k)$ of the cost function. Moreover, if $l$ is minimised, then the inner minimisation of (20) can be removed. So with the vectors

$$v(k) = [u_F(k) \quad u_B(k) \quad u_{PV}(k) \quad \delta_F(k) \quad \chi_F(k) \quad \chi_B(k)]^\top,$$
$$z^\sigma(k) = [p_F^\sigma(k) \quad p_B^\sigma(k) \quad p_{PV}^\sigma(k) \quad \delta_{PV}^\sigma(k) \quad \mu^\sigma(k)]^\top,$$

and decision variables

$$\mathbf{v} = [v(k+j|k)]_{j=0}^{J-1}, \quad \mathbf{x}^\sigma = [x^\sigma(k+j|k)]_{j=0}^J,$$
$$\mathbf{z}^\sigma = [z(k+j|k)]_{j=0}^{J-1}, \quad \mathbf{t} = [t(k+j|k)]_{j=0}^{J-1},$$

the mixed-integer-bilinear MPC problem reads as follows.

*Problem* 1 (Minimax MPC).

$$\min_{\mathbf{v}, \mathbf{x}^\sigma, \mathbf{z}^\sigma, \mathbf{t}} \sum_{j=0}^{J-1} t(k+j|k) \cdot \gamma^{j+1}$$

subject to (1)–(6), (10)–(13), $j = 0, \ldots, J-1$, $\sigma \in \{\text{high, low}\}$ with initial conditions, $x^{\text{high}}(k|k) = x^{\text{low}}(k|k) = x(k)$ and $\delta_F(k|k) = v(k-1)$.

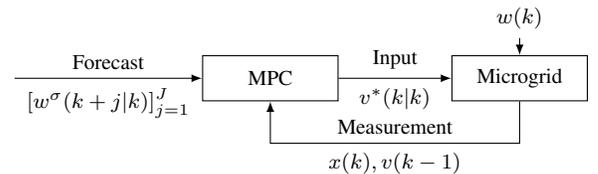

Fig. 2. MPC scheme for time instant k

TABLE II
SIMULATION PARAMETER

| Variable | Interval/Value | Variable | Value |
|---|---|---|---|
| $p_B$ | [-1, 1] | Prediction horizon $J$ | 6 |
| $p_{FC}$ | [0.2, 2] | Simulation duration | 48:00 h |
| $p_{PV}$ | [0, 4] | Sample time | 00:30 h |
| $\mu$ | [-0.314, 0.314] | $\chi$ | [0.314, 3.141] |
| $x$ | [0.2, 3] | Base pu | 10 kW |
| $c''_{PV}, c''_B$ | 1, 0.1 | $c_{Fru}, c'_{Fru}$ | 0.13, 1.56 |
| $c_{Fsw}, \gamma$ | 0.9, 0.2 | $\delta_{FC}(0), x(0)$ | 1, 0.3 |

Forecast bounds up to $J$ and initial conditions are fed in to the MPC, where Problem 1 is solved in a receding horizon manner. The resulting optimal setpoints for the first prediction step are then applied to the MG, which outputs the next state and current control input as next initial conditions.

*Remark* 1. The MPC includes two power trajectories, but only one optimal control input. Applying it to the MG model, yields power values that lie between the MPC power trajectories.

### III. CASE STUDY

To run closed-loop simulations with the MPC scheme, robust forecast intervals are required. These are obtained by generating a set of $N \in \mathbb{N}$ independent forecast scenarios, from which the upper and lower bounds of the distribution can be derived. Given a scenario $\hat{w}_i^{(l)}(k+j|k)$, where $l \in \mathbb{N}_{[1,N]}$, the bounds of unit $i$ can be obtained via

$$w_i^{high}(k+j|k) = \max_{l \in \mathbb{N}_{[1,N]}} (\hat{w}_i^{(l)}(k+j|k)), \quad (21a)$$

$$w_i^{low}(k+j|k) = \min_{l \in \mathbb{N}_{[1,N]}} (\hat{w}_i^{(l)}(k+j|k)). \quad (21b)$$

In the case study, 100 forecast scenarios are generated using an ARIMA(1,0,0)(0,1,0)$_{48}$ model for irradiance and an ARIMA(0,0,0)(0,1,0)$_{48}$ model for load forecast, trained with data from [18] and [19], respectively.

The MPC scheme is implemented in two distinct configurations. In the first one, the gains are $\chi_F(k) = \chi_B(k) = 0.5, \forall k$. In the second configuration, they are decision variables in (10) and (11). Furthermore, the frequency is limited by $\mu_{min}$ and $\mu_{max}$ to a $\Delta f$ of $\pm$ 0.5 Hz. The parameters are listed in Table II.

To highlight the advantages of the adaptive gain configuration over a fixed one, closed-loop simulations are performed. The simulations are conducted with an Intel® Core™ i7-10510U processor @ 1.80 GHz over a period of 48 h with a sample time of 30 min. In Fig. 3, the simulation results are illustrated. The resulting trajectories exhibit repeated peaks in load and PV infeed. These cycles can be classified into three distinct regions: zero (I), low (II) and high (III) PV infeed.

The simulations start at 00:00 h, with an initial state-of-charge (SoC) of 0.3 puh. As no PV infeed is available at this time, the remaining storage energy is consumed, leading to low SoC values. In these regions, although SoC levels remain low, the adaptive gain configuration maintains a small energy reserve, keeping the battery as the primary source for deviation handling, except when the lower SoC threshold is reached, as observed in Fig. 3. Furthermore, $\mu$ is mainly positive, as the adaptive gains enable less conservative grid-forming setpoints.

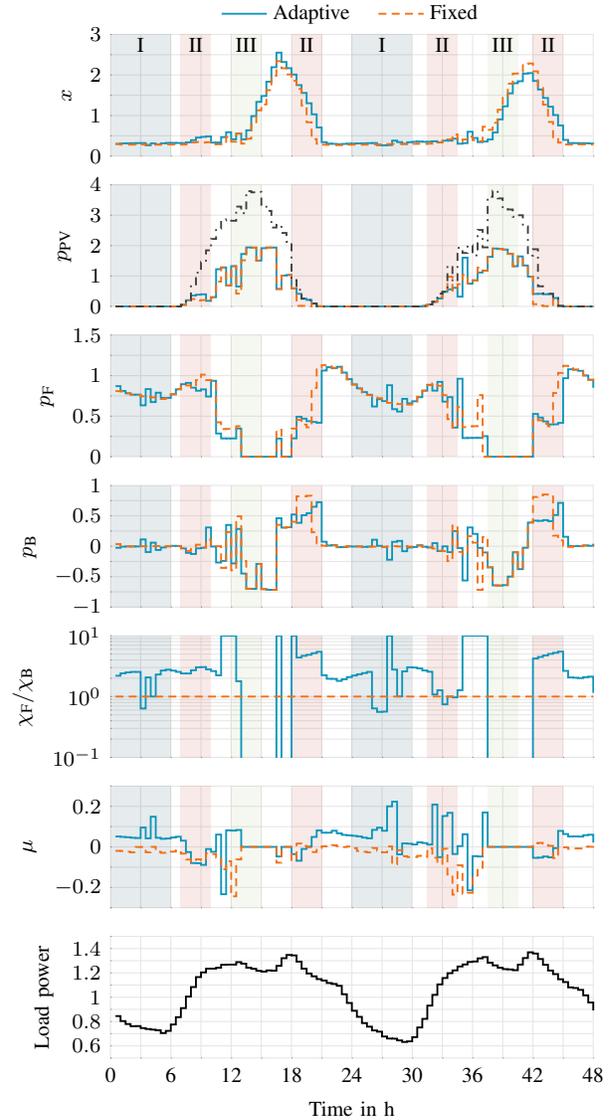

Fig. 3. Simulation results for adaptive and fixed configurations in pu. The regions highlight, where renewable infeed is zero (I), low (II) and high (III).

The second kind of regions corresponds cases, where the available PV infeed initially rises from or declines to zero (see Fig. 3). In the fixed configuration, (8) and (13) enforce a tight alignment between power and setpoint trajectories, which is realized through conservative PV curtailment and increased FC power. On the other hand, the adaptive gains in (10) and (11) relax the alignment between power and setpoints, allowing for larger, distinct offsets. Additionally, in these regions the battery predominantly handles fluctuations and $\mu$ is negative due to less conservative battery setpoints. This changes when large infeed deviations occur, requiring additional FC support.

Regions III correspond to periods where the available PV infeed reaches its peak. In the adaptive configuration, as sufficient energy has been stored, higher PV setpoints are possible and consequently at some times more battery charging is allowed. Moreover, in both cases the FC is shutdown leading to a $\chi_\mathrm{F}$ of 0. The adaptive approach yields 7.5 % more renewable infeed and a 6 % reduction in FC utilization over the entire simulation duration. However, the improvement entails increased complexity, with the mean solve time being more than four times higher due to the bilinear formulation. Nevertheless, due to a sampling time of 30 min, the computation times remain sufficiently fast.

TABLE III
Results

|  | Adaptive | Fixed |
|---|---|---|
| Cumulative PV power | 24.18 puh | 22.489 puh |
| Cumulative FC power | 26.734 puh | 28.416 puh |
| Mean/Max solve time | 0.058 s / 0.584 s | 0.013 s / 0.1079 s |

## IV. Conclusion

A bilinear formulation of adaptive power sharing is derived in this work, integrated in a robust minmax MPC scheme, and run in closed-loop simulations. Aiming to highlight the advantages of said adaptive formulation, it is compared to the same robust MPC with fixed gains. The adaptive configuration increases the PV infeed particularly at the lower available infeed values due to the increased flexibility introduced to the robust formulation. Future work will include a larger MG to test scalability, experimental validation and the integration of a more detailed, dynamical model of lower control layers.

## Acknowledgments

This research paper is funded by dtec.bw–Digitalization and Technology Research Center of the Bundeswehr–which we gratefully acknowledge. dtec.bw is funded by the European Union–NextGenerationEU.